## ASTRONOMY

# Magnetar formation through a convective dynamo in protoneutron stars

Raphaël Raynaud[1]*, Jérôme Guilet[1], Hans-Thomas Janka[2], Thomas Gastine[3]

The release of spin-down energy by a magnetar is a promising scenario to power several classes of extreme explosive transients. However, it lacks a firm basis because magnetar formation still represents a theoretical challenge. Using the first three-dimensional simulations of a convective dynamo based on a protoneutron star interior model, we demonstrate that the required dipolar magnetic field can be consistently generated for sufficiently fast rotation rates. The dynamo instability saturates in the magnetostrophic regime with the magnetic energy exceeding the kinetic energy by a factor of up to 10. Our results are compatible with the observational constraints on galactic magnetar field strength and provide strong theoretical support for millisecond protomagnetar models of gamma-ray burst and superluminous supernova central engines.

## INTRODUCTION

Magnetars are isolated, slowly rotating neutron stars characterized by a variable x-ray activity, which is thought to be powered by the dissipation of strong magnetic fields (*1*). Their measured spin-down is related to a dipolar surface magnetic field ranging from $10^{14}$ to $10^{15}$ G.

Although magnetic flux conservation during stellar collapse is commonly invoked to explain pulsar magnetism, it tends to fall short for these stronger field strengths (*2*). Furthermore, a strong magnetic field brakes stellar rotation (*3*) such that a fossil field is probably not compatible with the fast rotation needed for millisecond magnetar central engines. Alternative scenarios preferentially rely on the magnetic field amplification by a turbulent dynamo, which could be triggered by the magnetorotational instability (*4*, *5*) or by convection (*6*). The former may grow in the differentially rotating outer stable layers, while strong convective motions develop deeper inside the protoneutron star (PNS) during the first 10 s following the core collapse of a massive star. However, the generation of a dipolar component compatible with observational constraints has never been demonstrated through direct numerical simulations.

To that end, we build a self-consistent, three-dimensional (3D) PNS model solving the nonlinear, magneto-hydrodynamic anelastic equations governing the flow of an electrically conducting fluid undergoing thermal convection in a rotating spherical shell. This sound-proof approximation, routinely used to model planetary and stellar dynamos, is justified in the convective zone of a PNS because the Mach number is low (Ma ≃ 0.02 to 0.05). The equations are integrated in time with the benchmarked pseudo-spectral code MagIC (*7*, *8*).

## RESULTS

The anelastic background state is assumed to be steady and isentropic and matches the structure of the PNS convective zone 0.2 s after bounce given by a 1D simulation of a core-collapse supernova of a 27 $M_{Sun}$ progenitor. At this time, the convective zone extends from $r_i$ = 12.5 km to $r_o$ = 25 km, 15 km below the PNS surface (fig. S1). We take into account the physical diffusion processes to compute the fluid viscosity ν, thermal conductivity κ, and magnetic diffusivity η consistent with the PNS structure (fig. S2). Convection is driven by a fixed heat flux at the boundaries, which we estimate using the 1D model to $\Phi_o \sim 2 \times 10^{52}$ erg/s. We apply stress-free boundary conditions for the velocity field for which the total angular momentum of the system is conserved. Because of the extremely large value of the conductivity, the magnetic boundary condition is set by assuming that the material outside the convective zone is a perfect electrical conductor. For the sake of comparison, we also carried out simulations with an outer pseudo-vacuum boundary condition. Nevertheless, the need for local boundary conditions necessarily introduces some undesired discontinuities, and a stronger validation of the magnetar central engine scenario would require a direct and simultaneous modeling of the outer 10 km below the PNS surface. On the other hand, the restriction of the computational domain to the convective zone allowed us to perform a parameter study consisting of 31 simulations in which we vary the PNS rotation rate Ω and the magnetic diffusivity by changing, respectively, the Ekman number E = $\nu_o/(\Omega d^2)$, where $d = r_o - r_i$ is the shell gap, and the magnetic Prandtl number Pm = $\nu_o/\eta_o$ (table S1). We focus on the regime of fast rotation corresponding to periods of a few milliseconds, which translates in low to moderate Rossby numbers Ro = $U/(\Omega d) \in [10^{-2}, 1]$, where $U$ is the root mean square fluid velocity.

Figure 1 is representative of the time evolution of dynamos obtained with short rotation periods. The kinematic phase is characterized by an oscillatory mode, which saturates below equipartition, which is typical of an αΩ dynamo driven by a large-scale differential rotation. More unexpected is the secondary growth following this first plateau, which results in a much stronger field dominated by its axisymmetric toroidal component. In our set of simulations, the transition takes from 1 to 5 s, which is longer than the kinematic growth but still shorter than the duration of the convective phase (*9*); in any case, the dynamo saturation is independent of the initial magnetic field, even for initial fields as low as $10^6$ G, which suggests that the bifurcation is supercritical. The changes in the field and velocity structures are illustrated by the insets in Fig. 1 and the 3D renderings of Fig. 2. In the saturated state, the strong toroidal magnetic field breaks the characteristic columnar structure of rotating convection (Fig. 2A), leading to the concentration of the most vigorous radial flows inside the tangent cylinder (Fig. 2B). We also note that in most simulations, the entropy field displays an equatorial symmetry breaking similar to what has been reported for fixed flux Boussinesq convection (*10*). This hemispheric asymmetry may be related to the emergence of the "Lepton-number emission self-sustained asymmetry" (LESA) observed in nonmagnetized core-collapse supernova simulations (*11*), suggesting that LESA may also take place in fast rotating, strongly magnetized PNS.

[1]AIM, CEA, CNRS, Université Paris-Saclay, Université Paris Diderot, Sorbonne Paris Cité, F-91191 Gif-sur-Yvette, France. [2]Max-Planck-Institut für Astrophysik, Karl-Schwarzschild-Str. 1, D-85748 Garching, Germany. [3]Institut de Physique du Globe de Paris, Sorbonne Paris Cité, Université Paris-Diderot, UMR7154 CNRS, 1 rue Jussieu, 75005 Paris, France.
*Corresponding author. Email: raphael.raynaud@cea.fr

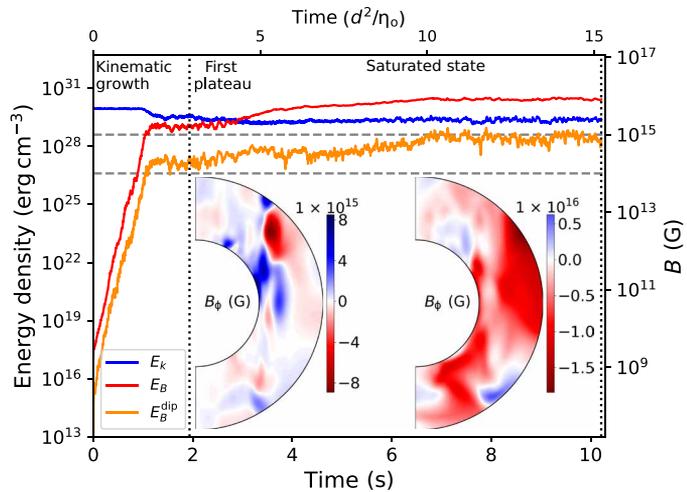

**Fig. 1. Time series of the densities of kinetic energy in the rotating frame (blue) and magnetic energy (red) for a fast-rotating model with rotation period $P = 2.1$ ms and Pm = 2.** The orange line represents the energy of the dipole component ($l = 1$ mode of the poloidal potential). The dashed gray horizontal lines show the fiducial range from $10^{14}$ to $10^{15}$ G for the intensity of the dipole field constrained by magnetar timing parameters. The insets show slices of the azimuthal magnetic field $B_\phi$ at times $t \sim 2$ s and 10 s indicated by vertical dotted lines. The upper x axis is labeled in units of magnetic diffusion time.

Our study demonstrates that this super-equipartition state is a new instance of a strong field dynamo characterized by the magnetostrophic balance between the Coriolis force $\tilde{\varrho}\Omega \mathbf{e}_z \times \mathbf{u}$ and the Lorentz force $\mu_0^{-1}(\nabla \times \mathbf{B}) \times \mathbf{B}$ (*12–15*). Figure 3 shows that our strong field solutions follow the expected scaling for the ratio of the magnetic and kinetic energies in the rotating frame, $E_B/E_K \propto \text{Ro}^{-1}$, which holds for rotation-dominated convective flows with Ro < 0.2. By contrast, weak field solutions are stable for slower rotation or when we use a pseudo-vacuum magnetic boundary condition, which artificially prevents currents outside of the convective zone, thereby forcing the toroidal magnetic field to vanish at the boundary. Since this unphysical constraint seems to impede a further growth of the toroidal field, we believe that the strong field branch will be achieved in future models including both the stable and unstable regions. It is noteworthy that the equipartition scaling often used to build quantitative models may actually underestimate the magnetic energy by a factor of 10 in rapidly rotating systems.

## DISCUSSION

This is the first numerical evidence that magnetar fields can be generated during the collapse of fast-rotating stellar cores, independently of their initial magnetization. Figure 4A demonstrates that PNS convective dynamos generate a large-scale dipole reaching up to $10^{15}$ G. Figure 4B shows that the toroidal magnetic field is always stronger than the dipole and reaches values as high as $10^{16}$ G. The ratio of poloidal to toroidal magnetic energy is shown in fig. S3; it may evolve during the relaxation process following the end of the convection within the PNS, but the nonzero magnetic helicity of this field configuration should prevent a substantial decay of the magnetic energy (*16*). In our scenario, the main requirement to produce a magnetar through a strong field dynamo is then that sufficient angular momentum is present in the progenitor star. From Fig. 4A, we deduce a minimum angular momentum $j > 4 \times 10^{15}$ cm$^2$/s to saturate on the strong branch (considering a 6-ms period and with a radius of 20 km).

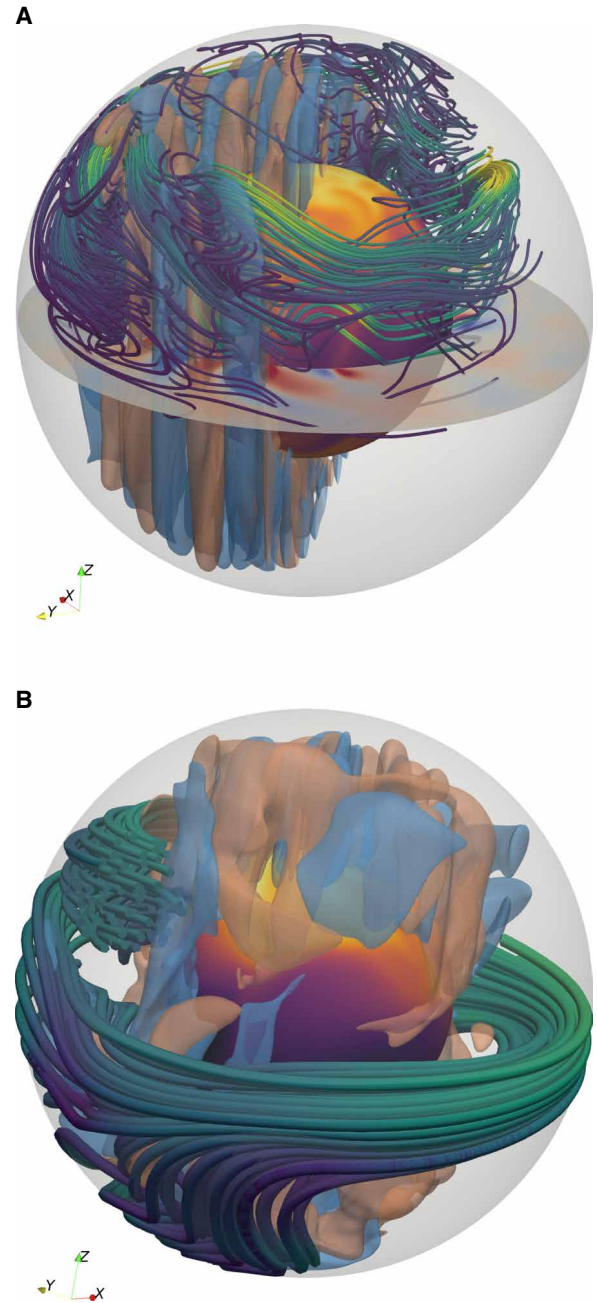

**Fig. 2. Three-dimensional rendering of the weak and strong field solutions.** Snapshots (**A**) and (**B**) correspond to the left and right insets in Fig. 1, respectively. Left-right snapshots correspond to the left-right insets in Fig. 1. Magnetic field lines are colored by the total field strength and the inner boundary by the entropy. Blue (red) isosurfaces of the radial velocity materializes the downflows (outflows). Radial velocities are of order $10^8$ cm/s.

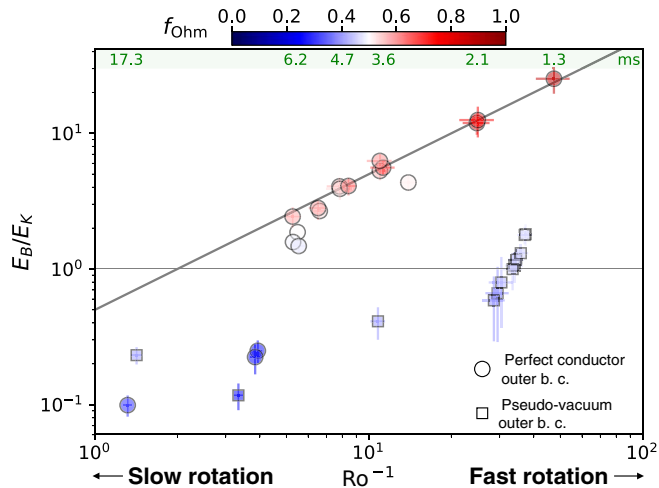

**Fig. 3. Ratio of the magnetic to the kinetic energy densities as a function of the inverse Rossby number.** The thick line shows the strong field best-fit scaling $\frac{E_B}{E_K} = 0.5\,\mathrm{Ro}^{-1}$, and the thin horizontal line corresponds to the equipartition scaling. Symbol color indicates $f_\mathrm{Ohm}$, the ratio of ohmic to total dissipation. The ohmic (resp. viscous) heating appears to be the dominant dissipation mechanism on the strong (resp. weak) field branch. The green banner indicates approximate rotation periods. The symbol shape indicates the type of the outer magnetic boundary condition (b. c.).

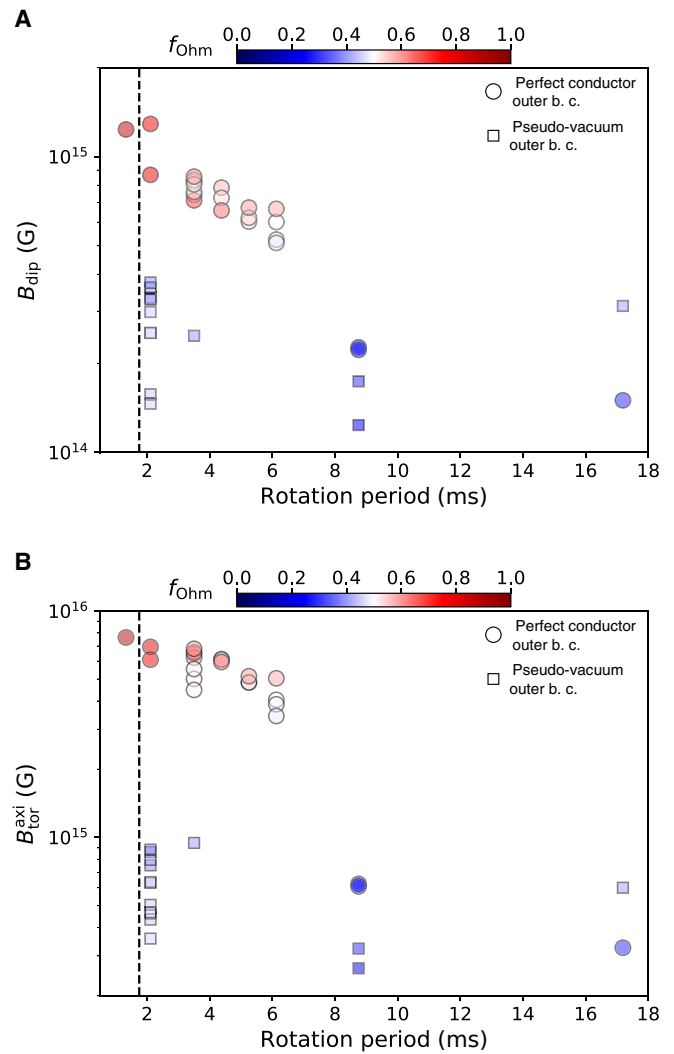

**Fig. 4. RMS values of the magnetic field as a function of the PNS rotation period.** (**A**) Intensity of the total dipole component. (**B**) Intensity of the toroidal axisymmetric component. The vertical dashed line shows the breakup rotation period $P_c = 2\pi\,(GM_{r<r_o}/r_o^3)^{-1/2} \sim 1.75$ ms. For a given rotation period, the vertical scatter is due to different Pm values.

These angular momentum values are consistent with gamma-ray burst (GRB) progenitor models from chemically homogeneous evolution—see fig. 2 in Woosley and Heger (*17*). The above value corresponds to a rotation period $P$ of 2.3 ms after contraction to a 12-km cold neutron star. This contraction will further amplify the magnetic field by a factor of 4 if the magnetic flux is conserved. We stress that the dynamo is tapping its energy from the gravitational energy through the work of the buoyancy force. The PNS rotation energy is not decreased and remains available to power a hypernova explosion and a relativistic jet. Our results therefore provide strong theoretical support for millisecond magnetar models of GRBs (*18*, *19*) by demonstrating that the right values of rotation period and dipole field can be consistently obtained with the assumed rotation rate. However, note that the magnetic field must rise sufficiently fast to the PNS surface to power these events; while our setup cannot describe this process, the magnetic fields we obtain lies well above the threshold for a rapid buoyant rise derived by Thompson and Murray (*20*). The magnetic energy exceeds $10^{50}$ erg in the form of toroidal field and provides an additional reservoir, which can be tapped on longer time scales to power x-ray flares (*21*) or repeating fast radio bursts (*22*).

While very fast rotation will lead to prompt strong dynamo and long GRBs, we argue that the more frequent case of (somewhat) slower rotation will lead to a delayed onset of the strong branch, which could explain superluminous supernovae (SLSNe) (*23*, *24*) and galactic magnetars. Over a few seconds, the PNS cooling and spin up due to contraction will contribute to decrease the Rossby number, providing even more favorable conditions to achieve a strong field dynamo at later times. Using a mixing-length argument to relate the energy flux to the convective velocity by $\Phi_o \propto R^2 \rho U^3 \propto M R^{-1} U^3$ (where $M$ and $R$ are the typical mass and radius of the convective zone), we obtain $\mathrm{Ro} \propto (\Phi_o R^4/M)^{1/3}/j$. The critical value $\mathrm{Ro} \sim 0.2$ translates into a specific angular momentum threshold $j > 9 \times 10^{14}$ cm$^2$/s, i.e., $P < 10$ ms for a strong field dynamo to appear at $t = 5$ s (using $\Phi_o = 4 \times 10^{51}$ erg/s and $R = 12$ km). Combining the above Rossby scaling and the magnetostrophic scaling demonstrated by our simulations, we express the magnetic flux as $R^2 B \propto (M^2 R \Phi_o)^{1/6} j^{1/2}$. This scaling predicts that dynamos acting at later time with lower values of $R$, $\Phi_o$, and $j$ will lead to weaker magnetic fields, with a dipolar component in the range of $[4, 8] \times 10^{14}$ G (using $2.5 < P < 10$ ms). It matches the magnetic fields deduced from the spin-down of galactic magnetars (*25*) and those obtained by fitting SLSNe with a magnetar model (*26*). In line with our reasoning, these values are lower than those derived from the modeling of long GRBs with millisecond magnetars (*27*).

In addition to generate a dipolar field compatible with magnetar timing parameters, our simulations predict an even larger toroidal component, which may provide an explanation for the free precession causing modulation of the x-ray emission (*28*) and the intriguing antiglitch phenomenon (*29*). They thus provide more realistic initial conditions to compute the long-term evolution of galactic magnetars (*30*). Furthermore, our scenario allows magnetar formation associated with supernovae with ordinary kinetic energies. Under the assumption that all the rotation kinetic energy ends up in the explosion kinetic energy, observational constraints from supernova remnants hosting a magnetar require an initial period longer than 5 ms (*31*). We then suggest that these galactic magnetars were born with periods between 5 and 10 ms.

The delay of the onset of a strong field dynamo for intermediate rotation periods is particularly interesting for SLSNe. SLSN magnetar models need the rotational kinetic energy to be released on time scales of days to weeks so that it can be efficiently radiated rather than being converted into the explosion kinetic energy (as for hypernovae). The delayed enhancement of the magnetic field after the explosion would lead to slower extraction of the rotation energy due to the lower density of its environment. The rotation energy is therefore kept for later times to power a SLSN.

Last, our results are also relevant to the context of binary neutron star mergers. The cooling of a neutron star remnant (if any) will likely lead to similar convective motions (*32*). Because of the rapid rotation, we predict that neutron stars formed in mergers have large-scale magnetic fields of magnetar strength generated by a strong field convective dynamo. This will be tested with dedicated numerical simulations and future multimessenger observations (*33*).

## MATERIAL AND METHODS
### 1D PNS model

The internal structure of the PNS with a baryonic (final gravitational) mass of 1.78 (1.59) $M_{Sun}$ is taken from a 1D core-collapse supernova simulation from Hüdepohl (*34*) (fig. S1). The calculations were performed with the code PROMETHEUS-VERTEX, which combines the hydrodynamics solver PROMETHEUS with the neutrino transport module VERTEX (*35*). VERTEX solves the energy-dependent moment equations of the three flavors of neutrinos and antineutrinos with the use of a variable Eddington factor closure and including an up-to-date set of neutrino interaction rates. The simulation used the nonrotating 27-$M_{Sun}$ progenitor s27.0 by Woosley *et al.* (*36*) and the high-density equation of state LS220 by Lattimer and Swesty (*37*). Our results are not very sensitive to slight changes in the progenitor structure (induced for instance by a varying rotation rate), since the 1D model only enters the definition of the anelastic reference state that approximates the background temperature and density radial profiles in the convective zone. The energy and lepton number transport by the convection inside the PNS is modeled with a mixing length treatment. We find that the Schwarzschild and Ledoux criteria lead to identical results when determining the location of the convective zone and neglect compositional effects due to lepton fraction gradients.

### Transport coefficients

In the PNS, energy is mainly transported by radiation consisting of neutrinos and antineutrinos of all flavors whose interaction with matter is determined by neutrino opacities. We assume that neutrinos are in thermal equilibrium with matter (their distribution therefore following the Fermi-Dirac statistics) and that their opacity increases with the square of the neutrino energy, $\chi_\nu = \frac{\chi_0}{E_0^2} E_\nu^2$. This gives the following expression for the gray energy flux

$$F = -\frac{1}{36} \frac{E_0^2}{\chi_0} \frac{k_B^2 T}{c^2 \hbar^3} \frac{\partial T}{\partial r} \quad (1)$$

where $c$, $\hbar$, and $k_B$ are the speed of light, the reduced Planck constant, and the Boltzmann constant, respectively. We sum all the different neutrino flavor contributions to define the fluid thermal conductivity $k$

$$k = \frac{1}{36} \frac{k_B^2 T}{c^2 \hbar^3} \left[ \left(\frac{E_0^2}{\chi_0}\right)_{\nu_e} + \left(\frac{E_0^2}{\chi_0}\right)_{\bar{\nu}_e} + 4 \left(\frac{E_0^2}{\chi_0}\right)_{\nu_x} \right] \quad (2)$$

where the subscript $_x$ refers to the other than electronic neutrino and antineutrino flavors. The different opacities are given by Janka (*38*)

$$\left(\frac{\chi_0}{E_0^2}\right)_{\nu_e} = \frac{\sigma_0 \varrho}{(m_e c^2)^2 m_u} \left[ \frac{5\alpha^2 + 1}{24} + \frac{3\alpha^2 + 1}{4} (1 - Y_e) \right] \quad (3)$$

$$\left(\frac{\chi_0}{E_0^2}\right)_{\bar{\nu}_e} = \frac{\sigma_0 \varrho}{(m_e c^2)^2 m_u} \left[ \frac{5\alpha^2 + 1}{24} + \frac{3\alpha^2 + 1}{4} Y_e \right] \quad (4)$$

$$\left(\frac{\chi_0}{E_0^2}\right)_{\nu_x} = \frac{\sigma_0 \varrho}{(m_e c^2)^2 m_u} \frac{5\alpha^2 + 1}{24} \quad (5)$$

In the above equations, $Y_e$ and $m_e$ are the electron fraction and mass, $m_u = 1.66 \times 10^{-24}$ g is the atomic mass unit, $\sigma_0 = 1.76 \times 10^{-44}$ cm$^2$, and $\alpha = -1.26$ is the charged-current axial-vector coupling constant in vacuum. Equation 5 describes the transport opacity for neutral-current scatterings on neutrons and protons. The opacities of electron neutrinos (antineutrinos) take into account an additional contribution describing the charged-current absorption on neutrons (protons). The thermal diffusivity $\kappa$ follows from the relation $\kappa = k/(\varrho c_p)$, where $c_p$ is the specific heat at constant pressure (fig. S2).

For the neutrino kinematic viscosity, we check the validity of the following approximation (*39*)

$$\nu = 1.2 \times 10^{10} \left(\frac{T}{10 \text{ MeV}}\right)^2 \left(\frac{\varrho}{10^{13} \text{ g cm}^{-3}}\right)^{-2} \text{ cm}^2/\text{s} \quad (6)$$

when compared to the results obtained with the second-order diffusion approximation.

Last, the electrical conductivity of degenerate, relativistic electrons scattering on nondegenerate protons is given by (*6*) $\sigma \sim \epsilon_F / (4\pi\hbar\alpha \ln \Lambda)$, where the symbols $\alpha$ and $\ln \Lambda \sim 1$ are the fine structure constant and the Coulomb logarithm, respectively. In the relativistic limit, the electron Fermi energies is $\epsilon_F = p_F c = \hbar k_F c$, with the Fermi impulsion $k_F^3 = 3\pi^2 n_e$ and the electron number densisty $n_e = \varrho Y_e/m_p$, where $m_p$ is the proton mass. Then, the magnetic diffusivity $\eta = c^2/(4\pi\sigma)$ scales like

$$\eta = 3.1 \times 10^{-5} \left(\frac{\varrho}{10^{14} \text{ g cm}^{-3}}\right)^{-1/3} \left(\frac{Y_e}{0.2}\right)^{-1/3} \text{ cm}^2/\text{s} \quad (7)$$

### Anelastic equations

We adopt the anelastic approximation to model PNS convection, which allows us to take into account the isentropic stratification of the convective zone while filtering out sound waves. We nondimensionalize the equations with the following units

$$t = \frac{d^2}{\nu_o} t^*, \quad \mathbf{u} = \frac{\nu_o}{d} \mathbf{u}^*, \quad \nabla = \frac{1}{d} \nabla^*, \quad p = \Omega \varrho_o \nu_o p^* \quad (8)$$

$$\mathbf{B} = \sqrt{\Omega \varrho_o \mu_0 \eta_o} \, \mathbf{B}^*, \quad S = d \frac{\partial S}{\partial r}\bigg|_{r_o} S^*, \quad \widetilde{T} = T_o \widetilde{T}^*, \quad \widetilde{\varrho} = \varrho_o \widetilde{\varrho}^* \quad (9)$$

where the tildes refer to background quantities. Dropping the superscript * of dimensionless variables, the Lantz-Braginsky-Roberts anelastic dynamo equations read in the PNS rotating frame (40 and references therein)

$$\nabla \cdot (\widetilde{\varrho} \mathbf{u}) = 0 \quad (10)$$

$$\frac{D\mathbf{u}}{Dt} = -\frac{1}{E} \nabla \left(\frac{p}{\widetilde{\varrho}}\right) - \frac{2}{E} \mathbf{e}_z \times \mathbf{u} - \frac{Ra}{Pr} \frac{d\widetilde{T}}{dr} S \mathbf{e}_r + \frac{1}{EPm} \frac{1}{\widetilde{\varrho}} (\nabla \times \mathbf{B}) \times \mathbf{B} + \mathbf{F}^\nu \quad (11)$$

$$\widetilde{\varrho} \widetilde{T} \frac{DS}{Dt} = \frac{1}{Pr} \nabla \cdot (\widetilde{\kappa} \widetilde{\varrho} \widetilde{T} \nabla S) + \frac{Pr}{Ra} \left(Q^\nu + \frac{\widetilde{\eta}}{Pm^2 E} (\nabla \times \mathbf{B})^2\right) \quad (12)$$

$$\frac{\partial \mathbf{B}}{\partial t} = \nabla \times (\mathbf{u} \times \mathbf{B}) - \frac{1}{Pm} \nabla \times (\widetilde{\eta} \nabla \times \mathbf{B}) \quad (13)$$

$$\nabla \cdot \mathbf{B} = 0 \quad (14)$$

In the above system, the viscous force $\mathbf{F}^\nu$ and viscous heating $Q^\nu$ are given by $F_i^\nu = \widetilde{\varrho}^{-1} \partial_j \sigma_{ij}$ and $Q^\nu = \partial_j u_i \sigma_{ij}$, where $\sigma_{ij} = 2 \widetilde{\varrho} \widetilde{\nu} (e_{ij} - e_{kk} \delta_{ij}/3)$ is the rate of strain tensor and $e_{ij} = (\partial_j u_i + \partial_i u_j)/2$ is the deformation tensor. Tensors are expressed with the Einstein summation convention and the Kronecker delta $\delta_{ij}$. The fraction of ohmic dissipation is defined by $f_{Ohm} = \langle Q^\eta \rangle / (\langle Q^\nu \rangle + \langle Q^\eta \rangle)$, where the angular brackets denote time and volume averaged quantities and $Q^\eta = \eta \mu_0^{-1} (\nabla \times \mathbf{B})^2$ is the Joule heating. The diffusivity profiles are normalized by their top values $(\widetilde{\eta}, \widetilde{\kappa}, \widetilde{\nu}) = (\eta/\eta_o, \kappa/\kappa_o, \nu/\nu_o)$. The background density $\widetilde{\varrho}$ and temperature $\widetilde{T}$ profiles are fitted from the PNS model described above with a fifth-order polynomial, which describes the profiles with a very good accuracy. The control parameters are the Ekman, Rayleigh, thermal, and magnetic Prandtl numbers defined, respectively, by

$$E = \frac{\nu_o}{\Omega d^2}, \quad Ra = \frac{T_o d^3 \frac{\partial S}{\partial r}\big|_{r_o}}{\nu_o \kappa_o}, \quad Pr = \frac{\nu_o}{\kappa_o}, \quad Pm = \frac{\nu_o}{\eta_o} \quad (15)$$

where the subscript designates quantities evaluated at the top of the convective zone.

To rescale the simulations, we estimate $\Phi_o = 4\pi r_o^2 \kappa_o \varrho_o T_o \frac{\partial S}{\partial r}\big|_{r_o} = 2 \times 10^{52}$ erg/s, $d = 12.5$ km, and $\varrho_o = 8.3 \times 10^{12}$ g cm$^{-3}$. We deduce the PNS rotation rate

$$\Omega = \left(\frac{\Phi_o}{4\pi r_o^2 \varrho_o d^3} \frac{Pr^2}{E^3 Ra}\right)^{1/3} \quad (16)$$

The values of the different diffusivities follow from the definitions in Eq. 15. As is usual in astrophysical fluid dynamics, the PNS parameter regime stands far beyond the reach of direct numerical simulations as far as the values of the diffusion coefficients are concerned. To limit the computational costs, we set the thermal Prandtl number $Pr = \nu_o/\kappa_o = 0.1$, while the magnetic Prandtl number mainly lies in the range Pm ∈ [2,10] (table S1). We stress that these moderate values ensue from the intrinsic limitations of the available computing power that will always prevent us to achieve with direct numerical simulations, a realistic scale separation between dynamical and resistive time scales. In consequence, the duration of the simulation is longer than the magnetic diffusion time (see Fig. 1), and the magnetic helicity is not conserved in our simulation, which is in contrast with ideal magneto-hydrodynamic (MHD) predictions (see fig. S4). However, it is still unclear whether the regime of high Pm will satisfy the conservation of the magnetic helicity because the small scales generated by turbulence can drive non-negligible dissipation even for extremely small resistivity.

### Numerical resolution and initial conditions
The grid resolution ranges from $(N_r, N_\theta, N_\phi) = (125,160,320)$ to $(257,512,1024)$. We have performed convergence tests to check that the results presented in this study are not sensitive to the simulation resolution. Futhermore, pseudo-spectral codes are less prone to numerical diffusion artefacts than other numerical methods. The empirical validation of convergence generally consists in checking that the kinetic and magnetic energy spectra display a decrease of more than two decades between the maximum and the smallest resolved scales. To be conservative, we have used a more stringent criterion with a minimum decrease of three decades (fig. S5).

The model displayed in Fig. 1 has been initialized with a seed magnetic field, restarting from an hydrodynamical simulation that reached a statistical steady state. However, to avoid the computation of the transient kinematic growth in each case, saturated dynamo solutions have also been used as initial conditions for different parameter values. We found that the dynamo saturation was independent of the initial magnetic field in the explored parameter range.

### SUPPLEMENTARY MATERIALS
Supplementary material for this article is available at http://advances.sciencemag.org/cgi/content/full/6/11/eaay2732/DC1

Fig. S1. Entropy per baryon and density profile inside the PNS 0.2s after bounce.
Fig. S2. Normalized diffusivity profiles as a function of radius.
Fig. S3. Ratio of the poloidal and toroidal magnetic energy.
Fig. S4. Time evolution of the magnetic helicity for a run that saturates on the strong field branch with $P = 2.1$ s and Pm = 2.
Fig. S5. Kinetic (blue) and magnetic (red) energy spectra.
Table S1. Overview of the numerical simulations carried out.

**Acknowledgments:** We thank R. Bollig, M. Bugli, T. Foglizzo, B. Gallet, D. Götz, and A. Reboul-Salze for the discussions and comments. We thank the anonymous referees for useful comments, which improved the quality of the paper. We thank the online CompOSE database (https://compose.obspm.fr). Numerical simulations have been carried out at the CINES on the Occigen supercomputer (DARI projects A0030410317 and A0050410317). **Funding:** R.R. and J.G. acknowledge support from the European Research Council (grant no. 715368, MagBURST). H.-T.J. is grateful for the support of the European Research Council (AdG no. 341157, COCO2CASA) and the Deutsche Forschungsgemeinschaft (grants SFB-1258 and EXC 2094). We thank the DIM ACAV and the CEA/IRFU for their financial support of the Alfvén cluster. J.G. acknowledges the support from the PHAROS COST Action CA16214 and the CHETEC COST Action CA16117. This work benefited from the PSI2 program "Gamma-ray bursts and supernovae: From the central engine to the observer". **Author contributions:** R.R. contributed to the conception of the project, the simulation code development, simulations, data analysis, visualization, and interpretation of results and wrote the manuscript. J.G. conceived the idea for the project and contributed to data analysis and interpretation of results and wrote the manuscript. H.-T.J. provided the PNS model and contributed to interpretation and preparation of the manuscript. T.G. contributed to simulation code development, data analysis, and interpretation of results and reviewed the manuscript. **Competing interests:** The authors declare that they have no competing interests. **Data and materials availability:** All data needed to evaluate the conclusions in this paper are present in the paper and/or Supplementary Materials. The code MagIC is freely available online at https://magic-sph.github.io. The simulation outputs are available upon request. Additional data related to this paper may be requested from the authors.

Submitted 4 June 2019
Accepted 17 December 2019
Published 13 March 2020
10.1126/sciadv.aay2732

**Citation:** R. Raynaud, J. Guilet, H.-T. Janka, T. Gastine, Magnetar formation through a convective dynamo in protoneutron stars. *Sci. Adv.* **6**, eaay2732 (2020).

# Science Advances

## Magnetar formation through a convective dynamo in protoneutron stars

Raphaël Raynaud, Jérôme Guilet, Hans-Thomas Janka and Thomas Gastine